\documentclass[twocolumn,showpacs,preprintnumbers,amsmath,amssymb]{revtex4}
\usepackage{graphicx}
\usepackage{dcolumn}
\usepackage{bm}

\begin{document}
\preprint{APS/123-QED}

\title{Nanoscale studies of domain wall motion in epitaxial ferroelectric thin
films}

\author{Patrycja Paruch}
 \email{patrycja.paruch@physics.unige.ch}
\author{Thierry Giamarchi}
\author{Thomas Tybell}
\altaffiliation[Now at ]{Department of Electronics and
Telecommunications, Norwegian University of Science and
Technology, N-7491 Trondheim, Norway.}
\author{Jean-Marc Triscone}
\affiliation{DPMC, University of Geneva, 24 Quai E. Ansermet, 1211
Geneva 4, Switzerland.}%

\date{\today}

\begin{abstract}
Atomic force microscopy was used to investigate ferroelectric
switching and nanoscale domain dynamics in epitaxial
Pb(Zr$_{0.2}$Ti$_{0.8}$)O$_3$ thin films. Measurements of the
writing time dependence of domain size reveal a two-step process
in which nucleation is followed by radial domain growth. During
this growth, the domain wall velocity exhibits a $v \propto
\exp{-(1/E)^{\mu}}$ dependence on the electric field,
characteristic of a creep process. The domain wall motion was
analyzed both in the context of stochastic nucleation in a
periodic potential as well as the canonical creep motion of an
elastic manifold in a disorder potential.  The dimensionality of
the films suggests that disorder is at the origin of the observed
domain wall creep. To investigate the effects of changing the
disorder in the films, defects were introduced during crystal
growth (a-axis inclusions) or by heavy ion irradiation, producing
films with planar and columnar defects, respectively. The presence
of these defects was found to significantly decrease the creep
exponent $\mu$, from 0.62 -- 0.69 to 0.38 -- 0.5 in the irradiated
films and 0.19 -- 0.31 in the films containing a-axis inclusions.
\end{abstract}


\maketitle

\section{Introduction}

Ferroelectric materials are widely used in modern technologies in
order to exploit their ferroelectric, piezoelectric and
pyroelectric properties. Standard devices generally use
microscopic patterned electrodes on ferroelectric ceramics, single
crystals or films to produce actuators, filters, resonators,
sensors and memories. However, the continuing demand for
miniaturization in acoustic applications \cite{larkin_saw_hf}, in
ultra-high density information storage
\cite{scott_memories,waser_memories}, and in
micro-electro-mechanical systems (MEMS), has made alternative
solutions necessary.  In terms of material growth, techniques like
rf-magnetron sputtering, pulsed laser deposition, and molecular
beam epitaxy with in-situ characterization \cite{ohnishi_MBE} have
been developed to precisely control the structure and thickness of
materials. Today, these techniques allow epitaxial thin films with
atomically flat surfaces and interfaces to be grown, both in
all-oxide structures and, more recently, on top of silicon
\cite{lin_PZT_Si}. Combined with these techniques, increased
understanding and control of oxide growth modes has permitted
self-assembly processes and nanostructuring in single crystals and
thin films \cite{lippmaa_nano}, a novel approach to
miniaturization. Low-temperature metallo-organic chemical vapor
deposition, more easily adaptable to industrial applications, has
also been actively researched to produce uniform films over large
areas with low fatigue \cite{asano_mocvd_caps}. As device
dimensions become reduced in extremely miniaturized systems, an
important issue is the ability to locally access and control
ferroelectricity. From its inception, there has been significant
interest in using atomic force microscopy (AFM)
\cite{hidaka_afm_FE,tybell_FE_films,gruverman_caps_AFM} for this
purpose, since it allows control over ferroelectric domain
structure at the nanoscopic scale required by ever smaller
systems.  For such systems, ferroelectric domain stability, domain
dynamics in the presence of electric fields or temperature
variations, and issues of domain growth are the main concerns.
Therefore, understanding the fundamental physics of ferroelectric
domains in thin films is of crucial importance.

\begin{figure}
 \includegraphics{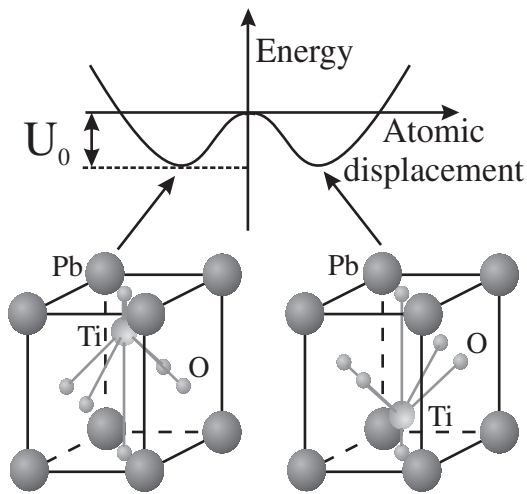}
 \caption{Schematic diagram of a Pb(Zr$_{0.2}$Ti$_{0.8}$O$_3$ unit
 cell in two degenerate ground states separated by an energy barrier $U_0$.
 The Pb and Ti/Zr ions are positively charged and the O ions are negatively
 charged, giving opposite polarization directions P$_{\rm DOWN}$ and P$_{\rm UP}$.}
 \label{FE}
\end{figure}

In this respect, it is useful to consider domain walls in
ferroelectric materials within the broader framework of elastic
disordered systems in the presence of an external force. Indeed,
ferroelectric materials are characterized by energetically
equivalent, degenerate ground states (tetragonal, in the case of
the perovskite materials used in this study), separated by an
energy barrier.  In a ferroelectric ground state the center of
gravity of positive charge in the unit cell is displaced relative
to the center of gravity of the negative charge, leading to a
stable remanent polarization reversible under an electric field,
as shown schematically in Fig.~\ref{FE}.
Regions of opposite polarization are separated by thin interfaces,
or domain walls. The application of an electric field asymmetrizes
the ferroelectric double well potential, favoring one polarization
state over the other by reducing the energy necessary to create a
nucleus with polarization anti-parallel to the field, thus
promoting domain wall motion. These domain walls can be considered
as elastic objects, whose surface tends to be minimized in order
to decrease the total wall energy. However, such elastic objects
may meander from an optimal flat configuration in order to take
advantage of particularly favorable regions of the potential
landscape, which can vary due to disorder in the ferroelectric
film, as well as the commensurate potential of the crystal lattice
itself. The behavior of such systems is thus governed by the
competition between elasticity and the effects of pinning.
Although at $T = 0$ the domain wall is pinned by the disorder
until a critical force $f_c$ is reached, at finite temperatures it
can be driven by forces below $f_c$, since barriers to motion,
however high, can always be passed via thermal activation. How the
domain wall moves when driven at small forces is of theoretical
and practical interest. A simple answer \cite{anderson_creep} is
motion governed by a linear response of the form $v \propto
e^{-\beta \Delta} f$ due to thermal activation above pinning
barriers $\Delta$, where $\beta$ is the inverse temperature
$1/{\rm k_B}T$ and $f \ll f_c$ is the driving force. However,
subsequent detailed analysis showed that both periodic
\cite{miller_nucleation_FE,blatter_vortex_review} and disordered
pinning potentials
\cite{ioffe_creep,nattermann_rfield_rbond,chauve_creep} can lead
to diverging barriers and thus to a nonlinear ``creep'' response,
in which the velocity is, for small temperatures and for forces
much smaller than the critical force, of the form $v \propto
\exp{(-\beta U_c (f_c/f)^\mu)}$. $U_c$ is a barrier height and
$\mu$ is an exponent characteristic of the mechanism responsible
for the creep. In a periodic potential creep occurs only for
interfaces of dimension $d=2$ (sheets), and $\mu = 1$
\cite{blatter_vortex_review}. For creep in a random potential $\mu
= (d - 2 + 2\zeta) / (2-\zeta)$, where $d$ is the dimensionality
of the interface and $\zeta$ its equilibrium roughness exponent
(see Section~\ref{sec:random}). Note that the nature of the
disorder only enters through the static roughness exponent
$\zeta$. Measurements of domain wall motion and equilibrium
roughness configuration in ultrathin magnetic films have verified
the creep law prediction for interfaces, with the measured
exponent $\mu = 0.25$ in good agreement with the expected
theoretical values for this system \cite{lemerle_FMDW}. In
periodic vortex systems, precise determination of the exponents
has been complicated by the multiple scales of the problem
\cite{blatter_vortex_review,kim_decoration_nbse}, although results
in agreement with the theoretical predictions of $\zeta =0$ have
been observed \cite{fuchs_creep_bglass}. High quality epitaxial
ferroelectric thin films are therefore another interesting model
system to study such phenomena, because their thickness and
crystalline quality can be precisely controlled, and because AFM
allows nanometer resolution of their domain configuration.

In this paper, we present a review of our studies of non-invasive,
local AFM control of ferroelectric polarization in epitaxial
Pb(Zr$_{0.2}$Ti$_{0.8}$)O$_3$ (PZT) thin films and its possible
applications. We will focus specifically on the behavior of domain
walls and its relation to domain growth and stability. We expand
on the discussion of our previous results demonstrating that
domain wall motion in thin films is a disorder-controlled creep
process. In particular, we discuss the differences between the
usual stochastic nucleation scenario proposed by Miller and
Weinreich to explain domain wall motion observed in bulk
ferroelectrics \cite{miller_nucleation_FE}, and the creep scenario
due to the competition between a disorder potential and the
elasticity of the wall. Finally, we report on new studies of
ferroelectric materials in which domain dynamics are measured in
the presence of macroscopic defects, both columnar, in the form of
heavy ion irradiation tracks, and planar, in the form of a-axis
inclusions in the c-axis oriented films. We find that these
defects, macroscopic in relation to the thickness of the domain
wall, nonetheless have a significant effect on domain wall
dynamics, lowering the values of the critical exponent for domain
wall creep from 0.62 -- 0.69 to 0.38 -- 0.5 in irradiated films,
and down to 0.19 -- 0.31 in films containing a-axis inclusions
\cite{PP_a_axis}.

\section{Thin film fabrication and AFM writing}

\begin{figure}
 \includegraphics{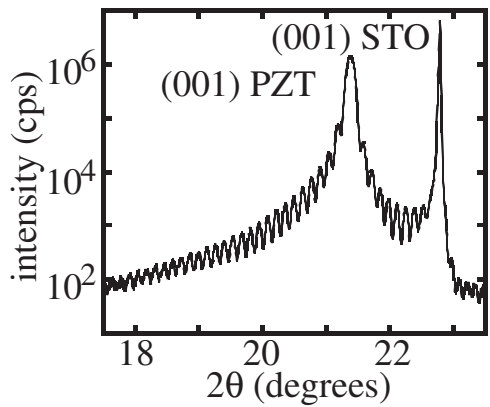}
 \caption{$\theta$--$2\theta$ scan of the PZT 001 peak in a 50 nm
 epitaxial thin film.  Multiple orders of finite size reflections
 allow precise measurement of film thickness.} \label{XRD}
\end{figure}

\begin{figure}
 \includegraphics{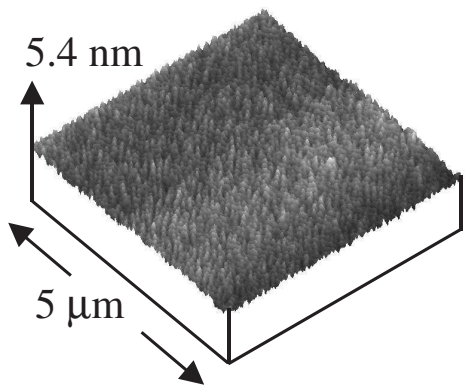}
 \caption{AFM topography of a 54 nm PZT film showing 0.22 nm rms
 roughness over a 5 $\times$ 5 $\mu$m$^2$ area.} \label{topo}
\end{figure}

The ferroelectric materials used in this study were c-axis
oriented PZT thin films, epitaxially grown by off-axis
rf-magnetron sputtering onto conductive (0.5 \% wt) Nb:SrTiO$_3$
(001) substrates in an Ar:O$_2$ flow at 180 mTorr, and at
substrate temperatures of $\sim$500$^{\circ}$C. Multiple samples
with thicknesses varying from 29.0 to 130.0 nm were grown.  X-ray
characterization of the films, such as the $\theta$--$2\theta$
diffractogram shown in Fig.~\ref{XRD}, revealed high crystalline
quality, with $\phi$ scans (not shown) confirming epitaxial
``cube-on-cube'' growth of the ferroelectric material on the
substrate.  The multiple orders of satellite reflections around
the principal 001 PZT peak are due to the finite size of the
sample, and allow us to precisely determine the thickness of the
films.
Measurements of sample topography showed uniform surfaces with
root-mean-square (rms) roughness of 0.2 -- 0.3 nm over 5 x 5
$\mu$m areas as shown in Fig.~\ref{topo}, where the vertical scale
is equal to 1 \% of the 54 nm thickness of the film itself. High
crystalline quality and smooth surfaces are advantageous for
effective writing and imaging of ferroelectric domains, since the
presence of morphological defects could perturb the tip-sample
interaction and thereby complicate the use of the films as model
systems for domain wall studies.

\begin{figure}
 \includegraphics{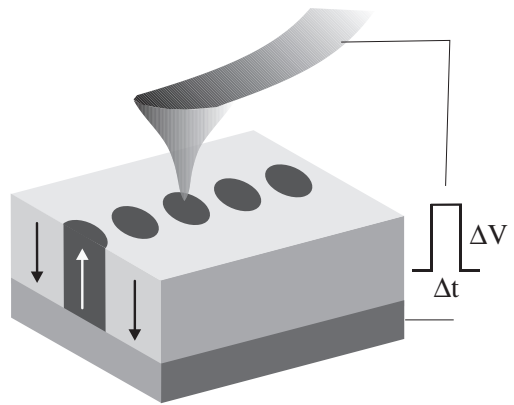}
 \caption{Schematic representation of local polarization switching
 in a ferroelectric thin film. The application of voltage pulses of
 magnitude $\Delta V$ and duration $\Delta t$ between a metallic
 AFM tip and the conducting substrate produces nanoscale circular
 domains penetrating through the thickness of the film.}
 \label{AFM}
\end{figure}

\begin{figure}
 \includegraphics{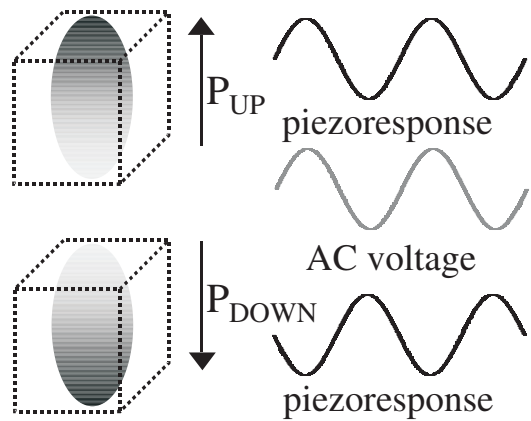}
 \caption{Schematic representation of the two antisymmetric
 polarization states in the tetragonal ferroelectric PZT unit cell.
 Applying a small AC voltage with an AFM tip excites a local
 piezoresponse, detected using a lock-in technique. The two
 polarization states (P$_{\rm UP}$ and P$_{\rm DOWN}$.) respond
 180$^{\circ}$ out of phase with each other, thus allowing a phase
 contrast image of the ferroelectric domains to be obtained.}
 \label{PFM}
\end{figure}

To locally control the ferroelectric polarization state, a
metallic AFM tip was used as a mobile top electrode, while the
conductive substrate functioned as a bottom electrode. Applying a
voltage between the tip and the substrate results in a local
electric field across the ferroelectric thin film, favoring the
polarization direction parallel to the field. This leads to
polarization switching if the field is high enough (or applied for
long times), as shown schematically in Fig.~\ref{AFM}. To image
the resulting domains the phase contrast of the local
piezoresponse of the film, excited by the application of a small
oscillating voltage via the AFM tip, (detailed in Fig.~\ref{PFM})
is measured by piezo-force microscopy (PFM). All measurements were
performed on commercially available {\it{Veeco Multimode}}
equipment.

\begin{figure}
 \includegraphics{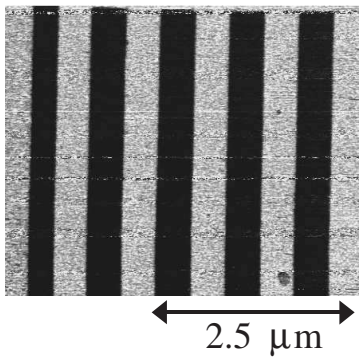}
 \caption{PFM image of AFM-written linear ferroelectric domains
 $925 \pm 15$ nm wide.  Similar domain structures have been used in
 a prototype surface acoustic wave device with GHz frequencies.}
 \label{SAW}
\end{figure}

Large areas, essentially limited only by the scan area of the AFM
and ultimately, the size of the sample itself, can be polarized by
applying a constant voltage to a tip scanning in contact with the
surface \cite{tybell_FE_films}. One can also alternate positive
and negative voltages applied for a fixed duration with respect to
the scanning speed of the sample to create arrays of lines with
opposite polarization, such as those shown in Fig.~\ref{SAW},
where the line width is $925 \pm 15$ nm. Such linear domain
structures have been used to develop a new type of interdigitated
transducer for a prototype high-frequency surface acoustic wave
device \cite{kumar_SAW}. Frequencies up to 3.4 GHz have been
demonstrated, but higher frequencies are in principle readily
accessible since AFM writing allows very small line widths. For
instance, using carbon nanotube bundles attached to a metallic AFM
tip to achieve a very high aspect ratio, we were able to write a
network of lines as small as 20-30 nm \cite{PP_AFM_nanotube}.
Small circular domains can also be created by applying short
voltage pulses to a stationary tip in contact with the sample, at
desired positions in a uniformly prepolarized area. As we have
previously shown, such domains can be written in arrays with
densities as high as $\sim$30Gbit/cm$^2$, with each domain (bit)
in such an array individually accessible and fully reversible
under subsequent voltage pulses \cite{PP_AFM_arrays}, an important
consideration from the point of view of dynamic non-volatile
memory applications.  Subsequently, Cho {\it{et al.}} have
demonstrated individual domains $\sim$15 nm in diameter in
ferroelectric single crystals which would give densities as high
as 0.5 Tbit/cm$^2$ \cite{cho_AFM_Tbit} when projected into a
standard array. By identifying the parameters controlling domain
size as the writing time and writing voltage (the duration  and
magnitude, respectively, of the voltage pulse applied to create
the domain), and the confinement of the electric field at the AFM
tip our studies address an active area of interest for
ferroelectric non-volatile memories: how to achieve the required
ultrahigh information densities.

\begin{figure}
 \includegraphics{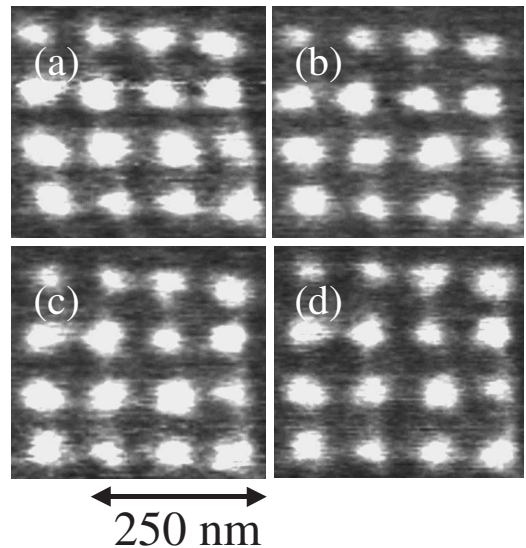}
 \caption{PFM images of the same array written with 500 ns, 12 V
 pulses, straight after writing (a), 1 week later (b), after 2
 weeks(c) and at the end of the experiment, a month later (d). The
 average domain radii are  29.2 nm, 27.0 nm, 27.2 nm and 27.7 nm
 respectively, unchanged within the $\pm$4 nm error of the
 experiment.} \label{stability}
\end{figure}

Two other key issues for information storage applications are the
minimum switching time and the stability of the resulting domains.
Using the AFM, we have written domains with pulses as short as 5
ns, with radii of $18.5 \pm 4$ nm.  This is not a fundamental
limit, however, but rather an experimental consideration related
to the minimum pulse time achieved by our pulse generator and the
RC characteristics of our system, which may lead to distortion of
the applied pulse at very short times. Recent studies by Li
{\it{et al.}} using a photoconductive switch and femtosecond laser
illumination suggest that the actual switching time for
ferroelectric domains may be as short as $\sim$70 -- 90 ps
\cite{li_laser_switching}. Our studies have also consistently
shown very high stability of AFM-written ferroelectric domains. An
array of 16 domains written with 500 ns pulses was imaged
immediately after writing (Fig.~\ref{stability}(a)), showing well
defined homogeneous domains with regular spacing and radii of 29.2
$\pm$4 nm.  The array was followed with measurements performed at
1 week (Fig.~\ref{stability}(b)), 2 weeks
(Fig.~\ref{stability}(c)) and 1 month (Fig.~\ref{stability}(d)),
with radii of 27.0 $\pm4$ nm, 27.2 $\pm$4 nm and 27.7 $\pm$4 nm
respectively.  No change, backswitching, or spontaneous
disappearance of the domains were observed. Additionally, the
arrays of lines used for surface acoustic wave devices were
measured over a period of 4 months, after lithographic patterning,
etching of areas outside the filter region, and the application of
voltage signals larger than those used for the piezoelectric
response in order to excite surface acoustic waves. Once again,
the domain structures remained completely stable \cite{kumar_SAW}.
Moreover, similar linear domains heated to 440$^{\circ}$C did not
show any change when imaged after subsequent cooling
\cite{ahn_undergrad}. It is therefore likely that pre-written
ferroelectric domain structures in thin films can subsequently be
incorporated into further processing steps up to, and possibly
well beyond this temperature. As shall be explained in more
detail, such high stability in zero applied field is in fact
inherent to the glassy behavior of an elastic disordered system.
\section{Experimental observation of domain wall creep}

The local control of ferrolectric polarization provided by AFM,
together with its nanometer resolution, also make it a powerful
tool for fundamental studies of domain dynamics and dependence on
writing parameters such as the writing time. As previously
reported \cite{PP_AFM_arrays}, we observed a strong dependence of
domain radius on the writing time for times longer than $\sim$20
$\mu$s. For shorter times, domain radii appeared to saturate at
$\sim$20 nm, a size we relate to the 25 -- 50 nm nominal radius of
curvature of the AFM tip used for these experiments. For each
writing time, 16- or 25-domain arrays were written with 12 V
pulses, and the domain radius was calculated by averaging over the
measured domain radii along their vertical and horizontal axes,
with a rms error of $\sim$10\%. All imaged domains appeared
homogeneous and well defined and no randomly nucleated domains
were observed. These data suggest a two-step switching process:
first, rapid nucleation and forward growth across the thickness of
the sample occur under the AFM tip; this event is followed by
slower radial motion of the domain wall outwards, perpendicular to
the polarization direction, in order to increase domain size.  We
analyzed this radial domain wall motion in the framework of a
pinned elastic system by comparing the velocity and the driving
force exerted on the wall, in our case due to the electric field
$E$ applied by the tip. We considered arrays written with
consecutive pulse durations, extracting the domain wall velocity
as $v=\frac{r(t_2)-r(t_1)}{t_2-t_1}$, the difference in domain
radii at the two subsequent writing times divided by the
difference in the writing times themselves. The electric field
distribution was obtained by modelling the tip as a charged
sphere, with radius $\alpha$ taken as equal to the domain
saturation size of 20 nm. Applying a voltage $V$ to the tip at the
surface of the ferroelectric film with dielectric constant
${\epsilon}$ produces a charge
$q=4{\pi}{\epsilon}{\epsilon}_0{\alpha}V$ on the model tip. Taking
into account both the effect of the film and the conductive
substrate, we are able to find the field $E_{\perp}(r,z)$ at any
point $(r,z)$ within the film. $r$ is the horizontal distance (in
the plane of the film) away from the center of the spherical tip,
and $z$ is the depth within the film (up to thickness $\lambda$)
from the center of the tip. In our experiments, the domain radii
remain comparable to the size of the tip, so further
simplification can be obtained by considering only the first order
of image charge reflections in the film and the substrate. Since
the voltage drop $V$ across the film is simply the integral of
this field over the film thickness $V = \int_{0}^{\alpha +
\lambda} E_{\perp}(r,z) dz$ we can define the average field across
the film $E(r) = \overline{E_{\perp}}(r)$, which shows a $1/r$
dependence in this first order approximation. As one moves further
away from the tip, a crossover to higher orders of $r$ dependence
in the denominator is expected for the field. This simplified
model shows reasonable agreement with a more accurate numerical
simulation of a hyperbolic tip in contact with a ferroelectric
film grown on a metallic substrate.  Here, a $1/r$ dependence of
the electric field is observed out to $r \sim 150$ nm, for a tip 1
$\mu$m high with ${\alpha}=20$ nm radius of curvature. Beyond
this, a cross over to $1/r^2$ and even steeper field decay occurs.

\begin{figure}
 \includegraphics{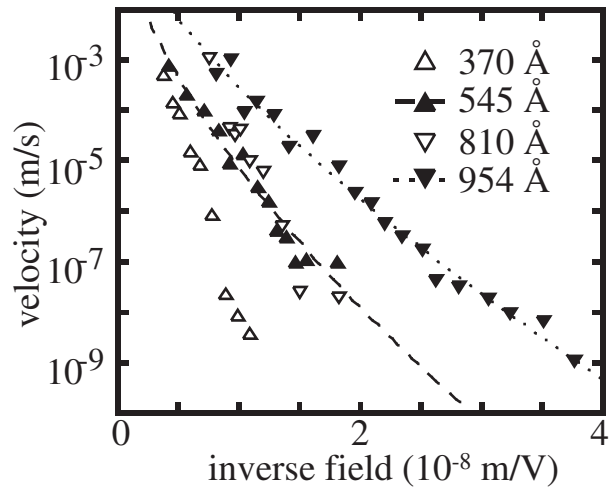}
 \caption{Domain wall speed as a function of the inverse applied
 electric field for 37.0, 54.5, 81.0 and 95.4 nm thick films.
 The data agree with the creep equation $v \sim
 \exp{[-\frac{R}{k_BT}\left(\frac{E_0}{E}\right)^\mu]}$ with $\mu$
 = 0.93, 0.62, 1.19, and 0.70 respectively. Fits of the data to
 $\log v = A(1/E)^\mu$ are shown for the 54.5 and 95.4 nm films.  }
 \label{arr_v_E}
\end{figure}

Although the field $E(r)$ is highly inhomogeneous at large length
scales, it can be taken as constant over the very small thickness
(on the order of a lattice spacing) of the ferroelectric domain
wall.  One can thus relate the velocity $v(r)$ of the domain wall
at a distance $r$ from the tip to a field
$E(r)=\frac{V{\alpha}}{r\lambda}$, where $r=\frac{r(t_1) +
r(t_2)}{2}$. An Arrhenius plot of the velocity against the inverse
field, shown in Fig.~\ref{arr_v_E}, reveals that our data are in
good agreement with a creep behavior
\begin{equation}
 v \sim \exp{-\frac{U_c}{{\rm k_B}T}\left(\frac{E_0}{E}\right)^\mu}
\end{equation}
over multiple decades of velocity, from 10$^{-3}$ to 10$^{-9}$
m/s, and for fields varying from 10$^7$ to 5 $\times$ 10$^8$ V/m
\footnote{We note that during AFM writing,the exact magnitude of
the effective field is difficult to quantify, because of a
possible gap between the film surface and the tip
\cite{hidaka_afm_FE}, and variations in tip shape. Local
piezoelectric hysteresis measurements on 120\AA - 800\AA\ thick
films show that the minimum switching field is $\sim$6--16 times
larger than the bulk coercive field, an effect that is not
observed with macroscopic electrodes on the similar films. The
effective field $E$ in the experiments is therefore presumably
$\sim$ one order of magnitude smaller that that calculated as
$E(r$) in this study. This has no effect on the exponent $\mu$
governing the exponential velocity dependence. Unless otherwise
noted, the values reported are the directly calculated ones, with
no further corrections.}. The values of the dynamical exponent
$\mu$ were found to be close to 1 for the original three samples
\cite{tybell_creep}, and closer to 0.7 for seven later samples --
grown and measured under the same conditions. These data were the
first indication that domain wall motion in ferroelectric thin
films was a creep process, and led us to investigate its
microscopic origins.

\section{Domain wall creep in a commensurate potential}

\begin{figure}
 \includegraphics{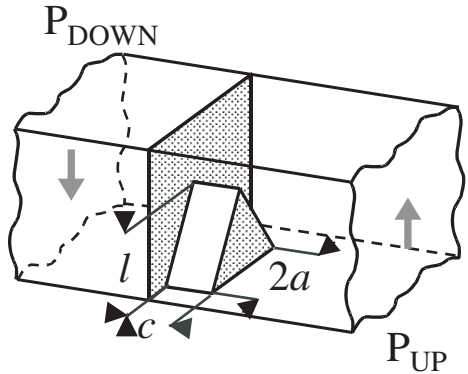}
 \caption{Schematic drawing of a triangular step domain on a
 180$^{\circ}$ domain wall, as described by Miller and Weinreich
 \cite{miller_nucleation_FE} The applied electric field $E$ is
 parallel to P$_{\rm DOWN}$.} \label{miller}
\end{figure}

Early studies \cite{merz_nonlinear_FE,fatuzzo_nonlinear_FE} of
domain growth carried out by optical and etching techniques on
bulk samples reported a non-linear electric field dependence of
the velocity $v \propto \exp{-1/E}$ known as Merz's law (with
implicit values of $\mu = 1$, if these results are to be
considered in the general framework of creep). At the time, a
phenomenological theory based on the stochastic nucleation of new
domains at existing domain boundaries was put forward by Miller
and Weinreich to explain the observed behavior
\cite{miller_nucleation_FE}. The wall moves forward due to the
formation of a nucleus as shown in Fig.~\ref{miller}. The energy
change due to the formation of a nucleus is
\begin{equation} \label{eq:miller}
\Delta F = - 2 P_s E V + \sigma_{w} A + U_{\rm depolarization}
\end{equation}
Nucleation would occur and the domain wall would move if the
energy gain, due to switching a volume $V$ of ferroelectric with
spontaneous polarization $P_s$ to the polarization state
energetically favorable with respect to the direction of the
applied field $E$, would balance the energy cost of extending the
surface $A$ of the domain wall, with a surface energy density of
$\sigma_w$, as well as the incurred depolarization energy cost
$U_{depolarization}$. In fact, this mechanism is identical
\footnote{In the absence of a depolarization field the nucleus is
isotropic. Taking into account the depolarization changes the
shape of the nucleus, but does not affect in an essential way the
physics leading to the creep process.} to the one of an elastic
manifold weakly driven in a periodic pinning potential (tilted
washboard potential, as described for example in
\cite{blatter_vortex_review}). The nucleus thickness ($c$ in
Fig.~\ref{miller}) is the distance between two mimima of the
periodic potential given by the lattice spacing of the
ferroelectric crystal.

For small electric fields ($E \rightarrow 0$) a large nucleus can
be expected since the energy gain due to the displacement of a
nucleus into the neighboring pinning valley grows with the volume
of the nucleus, while the energy cost essentially scales with its
surface. Since $V \sim L^d$ while $A \sim L^{d-1}$, where $L$ is
the extension of the nucleus, and $d$ the dimensionality of the
elastic interface, two cases occur. For a one-dimensional manifold
(string, $d=1$), the nucleus consists of two point-like kinks,
whose activation energy therefore always remains finite, and the
system exhibits a {\it linear} response under small driving
forces. For a two dimensional manifold, on the other hand,
minimizing (\ref{eq:miller}) gives $L^* \sim 1/E$, showing that
the size of the nucleus grows as the electric field decreases. The
energy barriers the nucleaus has to overcome thus grow as
$\Delta(E) \sim 1/E$, using (\ref{eq:miller}), giving a non-linear
response with $v \propto \exp{-1/E}$. The stochastic nucleation
scenario proposed by Miller and Weinreich can thus explain the
observed non-linear response of the domain wall {\it only if} the
domain wall itself is a two-dimensional surface embedded in a
three-dimensional crystal. This means that the dimensions of the
nucleus, at a given field $E$, have to be smaller than the
thickness of the system. Otherwise, the energy of the nucleus
saturates and the one dimensional wall case and a linear response
are recovered. It is also important to note that if the creep
consists of motion in a periodic potential the creep exponent is
constrained to be $\mu = 1$. As already mentioned, this particular
scenario is microscopically related to the intrinsic periodic
potential of the ferroelectric crystal itself acting to pin the
domain wall. The strength of this potential was calculated in
{\it{ab-initio}} studies of 180$^{\circ}$ domain walls in
PbTiO$_3$, showing that the wall energy varies from 132 mJ/m$^2$
to 169 mJ/m$^2$ depending on whether the domain wall is centered
on a Pb-O or Ti-O$_2$ plane in the crystal
\cite{meyer_periodic_FE_DW}. The influence of such a periodic
potential is due to the extreme thinness of the domain wall in
ferroelectric materials, in contrast to magnetic systems where the
domain wall is much larger than the atomic length scales
\cite{lemerle_FMDW}.

In order to test whether the observed creep behavior is indeed due
to the nucleation process, we calculated the size of the critical
nucleus, following the formulation derived by Miller and Weinreich
for the energetically most favorable dagger-shaped nucleus of
horizontal extension $a$, height $l$ and thickness $c$ forming at
an existing 180$^{\circ}$ domain wall, as shown in
Fig~\ref{miller} \cite{miller_nucleation_FE}, where $P_s$ is the
polarization, $b$ the in plane lattice constant, $\epsilon$ the
dielectric constant of PZT at ambient conditions, and $E$ the
applied electric field. The depolarization energy can be written
as $U_{\rm depolarization}=\frac{2{\sigma}_{p}ba^2}{l}$, with
$\sigma_p=(4P_{s}^{2}b \ln{(0.7358a/b)})/\epsilon$
\cite{miller_nucleation_FE}. By minimizing the free energy change
due to nucleation with respect to the dimensions of the nucleus
$a$ and $l$, with $c$ taken as equivalent to the lattice constant
$b$ (the distance between two minima in the periodic crystalline
potential), the size of the critical nucleus $a^\star$ and
$l^\star$, as well as the activation energy ${\Delta}F^\star$, can
be calculated as \cite{miller_nucleation_FE}:
\begin{eqnarray}
 a^\star &=& \frac{\sigma_w (\sigma_w + 2\sigma_p)}{P_{s}E(\sigma_w
 + 3\sigma_p)} \nonumber\\
 l^\star &=& \frac{\sigma_{w}^{1/2} (\sigma_w +
 2\sigma_p)}{P_{s}E(\sigma_w + 3\sigma_p)^{1/2}} \\
 {\Delta}F^\star &=& \frac{4b}{P_{s}E}\sigma_p(\sigma_{w} +
 2\sigma_{p}) \left(\frac{\sigma_w}{\sigma_w +
 3\sigma_p}\right)^{3/2} \nonumber
\end{eqnarray}
To compute the actual values, we take the standard parameters for
PZT ($P_s = 0.40$ C/m$^2$, $\epsilon = 100$, $b = 3.96$ {\AA} ),
and the {\it{ab-initio}} value for the domain wall energy
density\footnote{This value is computed for PbTiO$_3$. The
presence of Zr in PZT would lead to local variations of this
energy density.} $\sigma_w = 0.132$ mJ/m$^2$. In our case, the
applied electric field varied from $\sim 2$ to 20 MV/m (with the
factor 10 correction), depending on the thickness of the sample
used and the distance from the AFM tip, with the most intense
fields for thin films and small domains. Corresponding values of
$\sigma_p$ were between 1.6 and 0.9 J/m$^2$. Since $\sigma_p$ is
therefore greater than $\sigma_w$, following Miller and Weinreich,
the expressions for the critical values can be simplified to:
\begin{eqnarray}
 a^\star &=& \frac{2}{3}\frac{\sigma_w}{P_{s}E} \nonumber \\
 l^\star &=& \frac{2\sigma_{w}^{1/2} \sigma_{p}^{1/2}}{\sqrt{3}P_{s}E} \label{eq:scaling}\\
 {\Delta}F^\star &=&
 \frac{8b}{3\sqrt{3}P_{s}E}\sigma_{p}^{1/2}\sigma_{w}^{3/2}
 \nonumber
\end{eqnarray}
For the field range used, these equations would give critical
values of $a^\star \sim$ 12.5 -- 125 nm  and $l^\star \sim$ 53 --
710 nm. These results imply that for the given value of field, the
vertical size of the critical nucleus would exceed the thickness
of the film itself. This suggests that the films are in a
2-dimensional limit, with the domain walls acting as a
quasi-one-dimensional manifold, for which the Miller and Weinreich
stochastic nucleation model, or alternatively, weakly driven
motion through a periodic potential, could not explain the
non-linear response observed. Finally, the values of the dynamical
exponent we observe, generally not equal to one, are also a strong
indication that an alternative microscopic mechanism for the
observed creep process should be considered.

\section{Domain wall creep in a random potential} \label{sec:random}

In the alternative scenario of a canonical ``glassy'' system,  an
elastic manifold is weakly collectively pinned by the quenched
disorder potential present in the medium, with important
consequences for both its static and dynamic behavior. Since
disorder is always present in any realistic system it is a
feasible source of the observed domain wall creep in ferroelectric
thin films. Both for vortices in superconductors
\cite{blatter_vortex_review} and in magnetic systems
\cite{lemerle_FMDW}, disorder is clearly the driving mechanism
behind the observed creep behavior. In PZT, a solid solution of
20\% PbZrO$_3$ in 80\% PbTiO$_3$, the presence of Zr atoms is a
possible source of disorder, although preliminary studies of
domain wall dynamics in pure PbTiO$_3$ show similar behaviour to
that observed in PZT.  Vacancies and other defects in the lattice
structure are also likely sources of disorder.  In ferroelectric
films the presence of disorder would dominate domain wall
behaviour for both one and 2 dimensional walls at large scales.
However, given the thinness of the domain wall, we note that the
periodic potential of the crystal is also present in the problem.
In principle, ferroelectric films could therefore ultimately be
used to study the competition between a periodic potential and
disorder (see e.g. \cite{emig_commdisorder_long}).
\begin{figure}
 \includegraphics{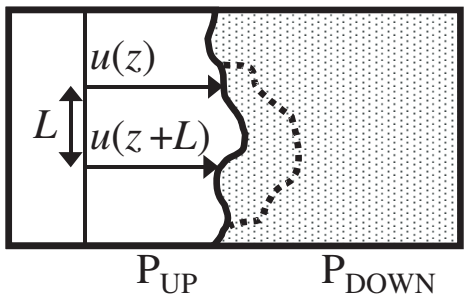}
 \caption{Domain wall as elastic manifold trapped in a random
 disorder potential.  In equilibrium, the domain wall exhibits a
 characteristic roughness, measured by the correlation function
 $B(z) = \overline{\langle[u(z+L) - u(z)]^2\rangle}$ of the displacements
 $u(z)$ from elastically ideal flat configuration with respect to
 the length L of domain wall. In the presence of a small driving
 force $f < f_c$ due to the applied electric field $E$, the domain
 wall will move to the next favorable configuration in the
 disorder potential, as shown by the dotted line, via a glassy
 creep motion. } \label{manifold}
\end{figure}

In order to analyze the effects of disorder on domain wall motion
let us again consider the energy for a segment of ferroelectric
domain wall of length $L$ displaced by $u(z)$ from the elastically
ideal flat configuration (where $z$ are the internal coordinates,
such that the total spatial dimension $d = z + 1$), as shown on
Fig.~\ref{manifold}. The energy scales as\footnote{There are
constants of order one, dependent on the dimension $d$, which have
been omitted from each term in the energy. These constants will
not affect the creep exponent $\mu$.}
\begin{equation} \label{eq:energy}
 U(u, L) = \sigma_{w}u^2 L^{d-2} - U_{\rm disorder}[u] - 2 P_{s}EL^d u
\end{equation}
where the first term describes the elastic energy contribution,
and is expressed for a local elasticity \footnote{Note that in
order to take into account the depolarization effects lengths
along the vertical axis have to be scaled by a factor
$(\sigma_p/\sigma_w)^{1/2}$, as in (\ref{eq:scaling}). Here $L$
denotes lengths perpendicular to the polarization direction.}. A
more accurate description of long range forces, such as dipolar
forces, modifies the elasticity and amounts to replacing $d$  by
$(3d-1)/2$ in the following formulas(see e.g.
\cite{emig_commdisorder_long} and ref. therein). The second term
is due to pinning by the disorder potential, and the third is the
energy due to the application of an external electric field.
$U_{\rm dis.}$ depends on the precise nature of the disorder. For
example for ``random bond'' disorder, equivalent to defects which
locally modify the ferroelectric double well potential depth, one
can model the disorder by a random potential acting at the
position of the interface and
\begin{equation}
 U_{\rm disorder} = \int d^dz V(u(z),z)
\end{equation}
Another form of disorder, the so called ``random field'', occurs
when defects locally asymmetrize the ferroelectric double well
potential, leading to a different form for $E_{\rm dis.}$. If the
disorder is weak, the central limit theorem allows its
approximation by a Gaussian random potential. The disorder is then
only characterized by its correlation length $r_f$ and the
strength of the random potential. In the absence of an external
electric field, the configuration of the domain wall results from
the competition between the elastic forces and the random
potential. This configuration can be characterized by measuring
the correlation function of relative displacements
\begin{equation}
 B(L)  = \overline{\langle [u(z+L)-u(z)]^2 \rangle} =
 \xi^2 \left(\frac{L}{L_c}\right)^{2\zeta}
\end{equation}
where $\langle \cdots \rangle$ denotes thermal averaging
(thermodynamic equilibrium) and $\overline{\cdots}$ denotes an
ensemble average over the realization of the disorder. In a
realistic experimental situation the ensemble average is performed
by averaging over all pairs of points separated by a distance $L$,
assuming that the system is self-averaging. $B(L)$ shows a
power-law growth with different exponents. For $r$ smaller than a
characteristic length, the Larkin length $L_c$
\cite{larkin_70,larkin_ovchinnikov_pinning}, $B(L)$ grows as $B(L)
\sim L^{4-d}$. Below this length there is no metastability and no
pinning of the elastic interface. Above the Larkin length, the
growth still follows a power law, but with an exponent $2\zeta$
($B(L) \sim L^{2\zeta}$) dependent on the nature of the disorder.
The Larkin length corresponds to the length for which the
displacements are of the order of the size of the interface or the
correlation length of the random potential \footnote{In this
simplified description we assume that the temperature is small
enough to neglect thermal effects.} $B(L_c) = \max(\xi,r_f)$. The
Larkin length is thus the smallest length at which the wall can be
weakly pinned, and above which it can adjust elastically to
optimize its local configuration \footnote{Above $L_c$, the domain
wall can also remain locally pinned on individual strong pinning
sites, but in the present discussion, only weak collective pinning
is considered.}. Above $R_c$ one can thus write
\begin{equation}
 B(L > L_c) = \max(\xi,r_f)^2 \left(\frac{L}{L_c}\right)^{2\zeta}
\end{equation}
The roughness exponent $\zeta$ is a function of the type of
disorder present in the film, and the dimensionality of the
manifold. For a line ($d=1$) and purely thermal fluctuations in
the absence of disorder, $\zeta = 0.5$. In a random bond scenario,
an exact value of $\zeta = 2/3$ has been calculated for a line
\cite{huse_DW_rb,kardar_comment,huse_response} and $\zeta \sim
3/5$ \cite{wolf_sim_RF_2d,forrest_roughening_2d} is expected for
the two-dimensional manifolds giving values of $\mu = 0.25$ and
$\mu \sim 0.5$ -- $0.6$, respectively, for the dynamical exponent
$\mu$ in these scenarios \footnote{For the random bond case, long
range dipolar forces would push in the two dimensional case the
exponent to $\mu \sim 0.66$ \cite{emig_commdisorder_long}}. We
note that random bond disorder exponents have been confirmed by
measurements of domain wall creep and roughness in an ultrathin
magnetic film \cite{lemerle_FMDW}. In a random field scenario, in
which defects locally asymmetrize the ferroelectric double well
potential, $\zeta = \frac{4 - d}{3}$ \cite{fisher_functional_rg},
giving $\mu=1$ for all dimensionalities of the manifold between 1
and 4.

\begin{figure}
 \includegraphics{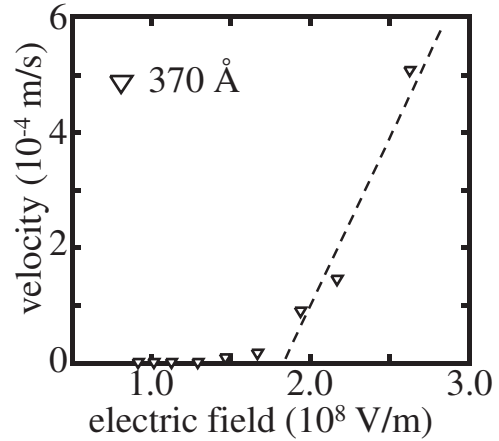}
 \caption{Domain wall velocity as a function of the applied
 electric field in a 37 nm film.  Extrapolating the linear
 behavior at high fields allows the critical field $E_c$ to be
 estimated as 100 MV/m.} \label{E_crit}
\end{figure}

$L_c$ is also the length scale at which pinning appears in the
system in the presence of a driving force. Using\footnote{We now
denote simply by $\xi$ the $\max(\xi,r_f)$.} (\ref{eq:energy}) for
$u \sim \xi$  and $L=L_c$ one can directly obtain\footnote{As
before the length here is the length perpendicular to the
polarization direction.} the critical field $E_c$
\begin{equation}
 E_c \simeq \frac{\sigma_{w}\xi}{P_s}\left(\frac{1}{L_c}\right)^2
\end{equation}
For driving forces above the critical force $f_c$, the interface
is unpinned even at zero temperatures since the force is large
enough to overcome the pinning barriers. For $f \ll f_c$, however,
the force is not large enough to overcome the barriers and the
motion proceeds by thermal activation. This is the creep regime,
leading to a small and non-linear response. In our case, a rough
estimate of the values of $E_c$ may thus be obtained by
extrapolating the linear behavior of the velocity, which occurs at
high field values.  Although we were unable to extend our
measurement significantly into this region, we can nonetheless at
least place a lower bound on the value of $E_c$ of 180 MV/m, as
indicated on Fig.~\ref{E_crit} for one of our thinner films, where
higher values of the field could be implemented. Taking $\xi$ to
be of the order of a unit cell, we can use the field data to
extract an approximate value of $L_c \sim 0.2$ nm, below the limit
of resolution of our measurement.

In the creep regime, we can rewrite (\ref{eq:energy}). For
simplicity we write formulas for the isotropic case. Using the
scaling $u \sim \xi (L/L_c)^\zeta$ one obtains
\begin{equation}
 E(u, L) = U_c\left(\frac{L}{L_c}\right)^{d - 2 + 2\zeta} - 2P_s E
 L_{c}^{d} \xi \left(\frac{L}{L_c}\right)^{d + \zeta}
\end{equation}
where $U_c = \sigma_w \xi^2 L_c^{d-2}$. Minimizing the energy with
respect to the external field $E$, we obtain the size of the
minimal nucleus capable of moving to a lower energy state as
\begin{equation} \label{eq:nucreep}
 L_{creep}/L_c = (f_c/f)^{1/(2-\zeta)}
\end{equation}
with $f = 2 P_s E$. The minimal barrier height to be passed by
thermal activation thus corresponds to the length $L^*$, leading
to a velocity of the form
\begin{equation}
 v \propto \exp(-\beta U_c (f_c/f)^{\frac{d-2+2\zeta}{2-\zeta}})
\end{equation}
if one assumes an Arrhenius law in passing the barriers. The very
slow (creep) response is due to the fact that for a small force
the system would have to rearrange large portions of the interface
to be able to find a new metastable state of low enough energy.
The barriers a domain wall must pass to make such a rearrangement
therefore diverge as the force goes to zero. This divergence could
inherently explain the very high stability of domain structures we
have observed in our films, in the limit of zero electric fields,
even at relatively high temperatures.

The expression (\ref{eq:nucreep}) gives the critical nucleus size
$L_{creep}$ as a function of the applied field $E$ and $L_c$. We
note that this expression is independent of the dimensionality of
the film, and that the applied and critical fields are present as
a ratio, thus removing the uncertainty associated with the
correction of the field in the AFM tip-ferroelectric thin film
configuration. As for the case of the periodic potential, these
expressions are valid if the size of the nucleus is smaller than
the thickness of the sample. Otherwise one of the dimensions of
the nucleus should be replaced by the thickness, transforming a
two-dimensional interface into a one-dimensional line. A crucial
difference between the periodic and the disordered cases is that
creep due to disorder can still exist in the one-dimensional
situation, contrary to the periodic case. Note that the question
of whether the films should be considered as one- or
two-dimensional depends on which mechanism controls the nucleus. A
film could thus be in the one-dimensional limit for the periodic
potential, thereby invalidating the periodic potential as a
possible origin for the creep process, and still be in the
two-dimensional limit for the disorder provided that the size of
the nucleus due to disorder remains smaller than the thickness of
the film. Although creep is still present in the one-dimensional
disordered case, the value of the exponent $\mu$ depends on the
dimension. Using the values for $E_c$ and $L_c$ we had obtained,
we can estimate the size of the critical nucleus for the creep
process and compare it with that found for the Miller-Weinreich
formulation. In our system, the applied field is a function of the
distance $r$ away from the tip center.  Using the largest possible
(random field) value of $\zeta$ we find $L_{creep}$ to vary
between 0.2 and 1 nm in the thinnest films (29.0 -- 51.0 nm), and
0.2 and 2.5 nm in the thickest films (95.0 -- 130.0 nm). Note that
in all cases $L_{creep} \sim 0.01r$, where $r$ is the radius of
the domain, so the approximation of a linear domain wall is quite
reasonable at that scale +\footnote{In the thinnest films, longer
writing times resulted in unstable and destructive interactions
between the tip and the sample. Therefore the larger domains
obtained for long writing times were only measured in the thicker
films.}.

Interestingly, the anomalously large size of the nucleus given by
the Miller-Weinreich model has previously been remarked on by
Landauer \cite{landauer_electrostatics}. Moreover, studies of the
piezoelectric effect and dielectric permittivity in PZT films
\cite{damjanovic_piezo,taylor_dielectric} have shown non-linear
features which cannot be described by a simple phenomenological
model, but which could be described by the pinning of domain walls
at randomly distributed pinning centers.\footnote{We note that the
studies referred to were carried out in thick (over 1 $\mu$m)
ceramic and sol-gel films, where the presence of multiple grain
boundaries and differently oriented domain walls provides a much
more complex disorder landscape compared to the epitaxially grown
films used for this study.} The scenario of weak pinning of
elastic interfaces in disordered media was also invoked as a
possible explanation for the dielectric dispersion observed in the
ferroelectric RbH$_2$PO$_4$ \cite{mueller_DW_pinning}. The present
study, investigating the static and dynamic behavior of individual
ferroelectric domains with nanoscale resolution therefore provides
an important clarification of this issue from the microscopic
point of view.

\section{Domain wall dynamics in the presence of artificial defects}

\begin{figure}
 \includegraphics{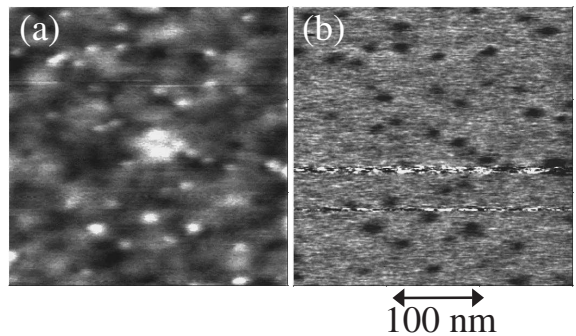}
 \caption{Topography (a) and PFM (b) measurements of a 95.4 nm
 sample with columnar defects created by heavy ion (Pb)
 irradiation.  On the film surface, the defects appear as white
 ``bumps'' in the topography, corresponding to the dark contrast
 regions in the PFM image, which we were unable to switch to the
 uniform polarization of the background.} \label{irr}
\end{figure}

\begin{figure}
 \includegraphics{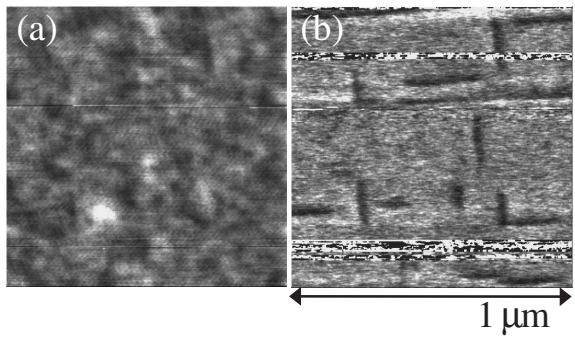}
 \caption{Topography (a) and PFM (b) measurements of a 108.9 nm
 sample with a-axis inclusions. These appear as criss-crossing
 lines on the surface, parallel to the a and b crystallographic
 axes of the film.  Since the direction of the polarization is in
 plane, these regions do not switch when a perpendicular electric
 field $E$ is applied across the film, once again appearing as dark
 contrast lines against a uniformly polarized background.}
 \label{axe_a}
\end{figure}

Another way to approach the question of ferroelectric domain wall
creep, and its microscopic origins, is to investigate the effects
of changing the disorder in the film.   We studied two different
types of defects: columnar tracks of amorphous material introduced
by heavy ion irradiation (carried out by C. Simon and A. Ruyter at
GANIL), and planar a-axis oriented inclusions introduced in
thicker films during growth. For the irradiated films, samples
51.0 to 116.0 nm thick were half-shielded by metallic plates
before irradiation, in order to directly compare domain dynamics
in non-irradiated and irradiated regions of the same sample. As
shown in the AFM topography scan in Fig.~\ref{irr}(a), the tracks
are 6 -- 13 nm in diameter at the point of entry, and appear as
small white ``bumps'', since the amorphous material is less dense
than the crystalline  form surrounding the tracks.  As shown in
the corresponding piezoresponse image (Fig.~\ref{irr}(b)), these
are exactly the regions which remain unaffected by attempts to
uniformly polarize the area by scanning with a constant negative
voltage, appearing as dots with dark phase contrast, as opposed to
the lighter phase contrast of the rest of the image. Thus,
irradiation defects can be distinguished from merely topographic
features like the large white ``bump'' in the middle of the
topographical scan, which switches under the application of an
electric field and does not appear as a corresponding dark area in
the piezoresponse image. The density of defects observed in this
image ($4.7 \times 10^{10}$ /cm$^2$) agrees reasonably with the
nominal irradiation density $4 \times 10^{10}$ /cm$^2$ used,
considering the small area ($300 \times 300$ nm$^2$) over which
the Poisson distribution is observed. To ensure continuous tracks
throughout the sample, with the ions lodging deep in the
substrate, measurements from other experiments on similar oxides
carried out at GANIL were used to determine irradiation energies
\cite{hardy_irr_high_Tc}.
To produce the planar a-axis inclusion,
we grew relatively thick films, slightly varying the temperature
from its optimal range, which generally results in the presence of
a-axis, as shown in the AFM topography (Fig.~\ref{axe_a}(a)). Once
again, the a-axis oriented regions do not respond to switching
attempts, as shown in the piezoresponse image in
Fig.~\ref{axe_a}(b), since the applied field is perpendicular to
the polarization axis. These defects are approximately 25 nm wide,
extending for hundreds of nanometers in the crystallographic a and
b axis directions, and inclined at 45$^{\circ}$ to the c-axis of
the film \cite{tybell_aaxis_tem}. In all films, 25 domain arrays
were written, and the domain dynamics extracted as described
previously \cite{tybell_creep}. In the a-axis films, domains were
written adjacent to, but not right on top of the a-axis
inclusions.

\begin{figure}
 \includegraphics{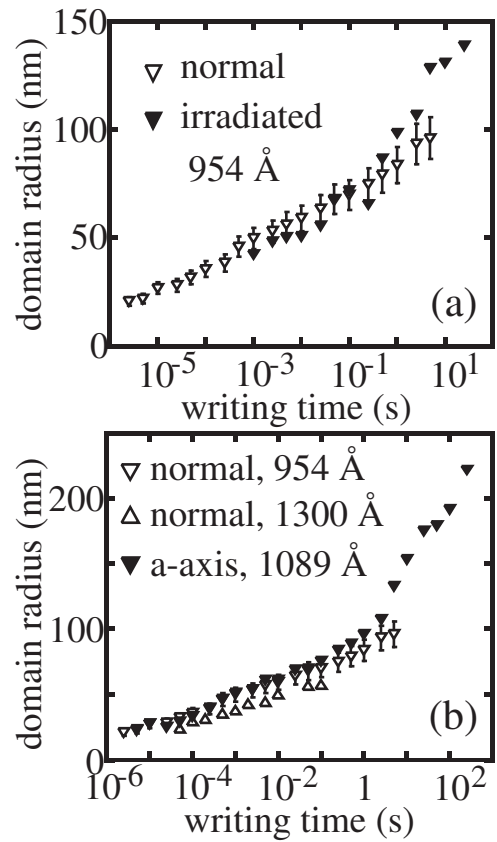}
 \caption{Domain radius as a function of the writing time for the
 95.4 nm half-irradiated film (a), and for the 108.9 nm film with
 a-axis inclusion (b).  In both cases, the domains written with
 longer writing times appear larger than in the non-irradiated or
 pure c-axis films included as a comparison.}
 \label{def_r_t}
\end{figure}

\begin{figure}
 \includegraphics{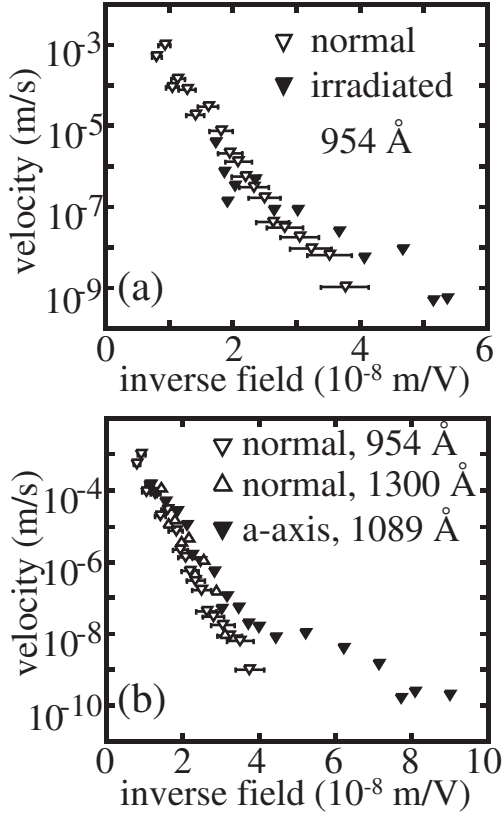}
 \caption{Domain wall speed as a function of the inverse applied
 field for the films shown in Fig.~\ref{def_r_t}.  In the presence
 of defects, the values of the creep exponent $\mu$ decrease from
 0.69 to 0.38 and from 0.69 and 0.78 to 0.31 in the half-irradiated
 film, and the pure c-axis films vs the film with a-axis inclusions, respectively.}
 \label{def_v_E}
\end{figure}

As shown in Fig.~\ref{def_r_t}(a) for a 95.4 nm film, we find that
domain sizes are comparable in both the irradiated and normal
regions of the sample.  However, for longer writing times, domains
in the irradiated part of the sample appear somewhat larger that
those written in the normal regions of the same sample.  When
comparing pure c-axis film 95.4 and 130.0 nm thick to a film of
similar thickness (108.9 nm) with a-axis inclusions
(Fig.~\ref{def_r_t}(b)) this increase in domain size for longer
writing times in the presence of defects is also observed.
The effect of the irradiation tracks and a-axis inclusions becomes
more evident when the Arrhenius plot of velocity vs the inverse
field is compared for the same films (Fig.~\ref{def_v_E}(a) and
(b)). We find a decrease in the values of the dynamical exponent
in the irradiated regions, as compared to the normal regions, from
a range of 0.62 -- 0.69 to one of 0.38 -- 0.5. This decrease is
even more marked for films containing a-axis inclusions, with
$\mu$ values of 0.31 and 0.19 \cite{PP_a_axis}. That this is not a
thickness effect can be seen by comparing the a-axis-containing
films to pure c-axis films of similar thickness, in which $\mu$
values of 0.69 and 0.78 (for 95.4 and 130.0 nm thick films,
respectively) are observed. It is interesting to note that both
these types of defects are large compared to the thickness of the
wall itself, occur at a relatively low density, and are either
columnar or planar, as opposed to point-like. The mechanism by
which they affect the dynamics of the domain wall is therefore
unlikely to be a direct weak collective pinning effect, and thus
the marked decrease observed in the dynamical exponent is somewhat
surprising. Possibly, the relaxation of strain in the film caused
by these macroscopic ``defects'' changes the density or pinning
force of the inherent, weak point defects present in the film,
thus affecting the disorder potential experienced by the domain
wall. Since a large amount of energy is dissipated within the
substrate by the heavy ion during irradiation, with possible
deformation as strain induced in the substrate
\cite{hensel_ybco_irr,konczykowski_pc}, this could result in
additional strain effects on the ferroelectric film.
 Alternatively, the presence of defects which penetrate through the
entire thickness of the film might act to make the wall more rigid
-- and therefore more one-dimensional. A more detailed
investigation of domain walls in the presence of these defects,
perhaps looking at their specific individual interaction, is
needed to ascertain the microscopic nature of this behavior.
\section{Conclusion}

Using the unprecedented control and precision provided by AFM, we
were able to study the growth of individual nanoscale
ferroelectric domains in epitaxial thin films, investigating both
the fundamental physics of domain wall motion, and its
consequences for possible AFM-ferroelectric applications. Our
studies demonstrate that domain wall motion in ferroelectric thin
films is a creep process in which $v \propto \exp(-\beta U_c
(E_c/E)^\mu)$, with a dynamical exponent $\mu$ between $\sim$ 0.7
and 1.0.  This process controls the lateral growth of domains in
low electric fields applied by an AFM tip.  A detailed analysis of
the possible microscopic origins of the observed domain wall creep
suggests that it is the result of competition between elastic
behavior and pinning in a disorder potential.  The reduced
dimensionality of our thin films compared to the size of the
critical nucleus precludes pinning in the commensurate potential
of the crystal itself as the mechanism for the non-linear field
dependence of the velocity. We have also observed that the
presence of artificially introduced defects, such as irradiation
tracks or a-axis inclusions, strongly decreases the dynamical
exponent for domain wall creep, from $\sim 0.6$ -- $0.7$ to $0.4$
-- $0.5$ and $\sim 0.31$. All the domains show high stability (up
to 4 months for the longest duration experiments), inherently
explained by the physics of a system in which elasticity and
pinning by a disorder potential compete, leading to glassy
behavior in the presence of low electric fields.  This high
stability is important for applications such as information
storage or surface acoustic wave devices based on ferroelectric
domain structures. Our studies have also identified the key
parameters controlling domain size as the writing time, the
applied voltage, and the confinement of the electric field, with
encouraging results for novel tip geometries using carbon nanotube
bundles.

\newpage

\begin{acknowledgements}
The authors would like to thank M. Dawber and C. H. Ahn for
enriching discussions. Special thanks to D. Chablaix for useful
technical developments and to S. Brown for careful reading of the
manuscript. This work was supported by the Swiss National Science
Foundation through the National Center of Competence in Research
``Materials with Novel Electronic Properties-MaNEP'' and Division
II.  Further support was provided by the New Energy and Industrial
Technology Development Organization (NEDO) and the European
Science Foundation (THIOX).
\end{acknowledgements}

\newpage
\bibliographystyle{prsty}

\end{document}